\documentclass[english,sort&compress]{iopart}
\usepackage[T1]{fontenc}
\usepackage[latin9]{inputenc}
\usepackage{babel}
\usepackage{varioref}
\usepackage{units}
\usepackage{amstext}
\usepackage{geometry}
\usepackage[numbers]{natbib}
\usepackage{xargs}[2008/03/08]

\makeatletter
\usepackage{iopams}
\usepackage{setstack}

\usepackage{hyperref}

\newcommand{\eqref}[1]{(\ref{#1})}

\makeatother

\begin{document}
\title[The Algebra of the Pseudo-Observables II]{The Algebra of the Pseudo-Observables II: The Measurement Problem}
\author{Edoardo Piparo\footnote{A.I.F. Associazione per l'Insegnamento della Fisica - Gruppo Storia della Fisica: \url{http://www.lfns.it/STORIA/index.php/it/chi-siamo}}}
\address{{\large I.I.S. ``Elsa Morante'', Viale Francesco Selmi 16, I-41049
Sassuolo (MO), Italy}}
\ead{{\large edoardo.piparo@istruzione.it}}
\begin{abstract}
In this second paper, we develop the full mathematical structure of
the algebra of the pseudo-observables, in order to solve the quantum
measurement problem. Quantum state vectors are recovered but as auxiliary
pseudo-observables storing the information acquired in a set of observations.
The whole process of measurement is deeply reanalyzed in the conclusive
section, evidencing original aspects. The relation of the theory with
some popular interpretations of Quantum Mechanics is also discussed,
showing that both Relational Quantum Mechanics and Quantum Bayesianism
may be regarded as compatible interpretations of the theory. A final
discussion on reality, tries to bring a new insight on it.
\end{abstract}

\maketitle
\global\long\def\ket#1{|#1\rangle}%

\global\long\def\bra#1{\langle#1|}%

\global\long\def\inprod#1#2{\left\langle #1\mid#2\right\rangle }%

\global\long\def\ideq{:=}%

\global\long\def\imp#1{#1_{\mathrm{I}}}%

\global\long\def\rp#1{#1_{\mathrm{R}}}%

\newcommandx\sspan[1][usedefault, addprefix=\global, 1=]{\mbox{span}#1}%

\section{Introduction}

\subsection{Measurement in Quantum Mechanics\label{subsec:Measurement-in-Quantum}}

Quantum measurement theory in standard textbooks is expressed in term
of the \emph{Copenhagen interpretation}, that can be summarized in
the following essential points:
\begin{enumerate}
\item A \emph{state vector} gives a \textbf{\emph{complete}} description
of the state of a physical system. It determines the probability distribution
of the measure outcomes of any observable quantity.
\item Our knowledge of the physical reality cannot be expressed but by means
of the language of the ``Classical Physics''. But the complete description
of the physical phenomena requires the use of contrasting classical
concepts, such as those of wave and particle. No contradiction, however,
arises from their use, since they describe phenomena that occur in
incompatible experimental situations (\emph{complementarity principle}).
\item The\emph{ uncertainty principle} states the impossibility of the simultaneous
existence of exactly defined values for two incompatible observables.
\item The observation results in a discontinuous change, that cannot be
described in terms of the time evolution equations, of the state vector
that causes a \emph{system state collapse} into one of the eigenstates
of the measured observable. This final state can be forecast only
\textbf{\emph{probabilistically}}. \textbf{The measuring apparatus
must be described classically}.
\end{enumerate}
The two key concepts of the interpretation are the \emph{completeness
of the quantum state}, as expressing \textbf{all that can be known}
about the physical system, and the the \emph{ontic probabilism}: the
need in the use of probability is not due to our lack of knowledge
but it's the measurement process itself to be intrinsically \emph{undeterministic}.

To be honest, this is \textbf{only one} interpretation of the Copenhagen
interpretation, since, according to Peres\citep{Peres2002}:
\begin{quotation}
``There seems to be at least as many different Copenhagen Interpretations
as people who use that term, probably there are more.''
\end{quotation}
Generally speaking, despite the wide-most popularity of such interpretation,
as witnessed by a poll executed by Schlosshauer \textit{et al.}\citep{Schlosshauer2013},
most physicists, as a matter of fact, have subscribed an instrumentalist
interpretation of quantum mechanics, a position, often equated with
eschewing all interpretations, summarized by the sentence \emph{``Shut
up and calculate!}'', using a slogan sometimes attributed to Paul
Dirac or Richard Feynman, but that seems to be due to David Mermin\citep{Mermin2004}.

One way or another, the quantum measurement presents several conceptual
problematic aspects, as it will be discussed in the next subsection.

\subsection{The measurement problem\label{subsec:The-measurement-problem}}

Pykacz, in his short survey on the main interpretations of Quantum
Mechanics\citep{Pykacz2015}, identifies, as drawbacks of the Copenhagen
interpretation, the artificial division of the physical world into
the quantum world and the classical world and the \textquotedblleft objectification
problem\textquotedblright , i.e., the problem of how the \textquotedblleft potential\textquotedblright{}
properties become \textquotedblleft actual\textquotedblright{} in
the course of a measurement. The latter is strictly tied to the so-called
\emph{measurement problem} in quantum mechanics, that is the problem
of how (or whether) wave function collapse occurs. The critical point
is that wave functions evolve deterministically according to the Schrödinger
equation, whereas the collapse is of a stocastical nature.

The problem gave rise to a plethora of interpretations of quantum
mechanics, all of them more or less unsatisfying for some aspect,
as evidenced, for instance, by the analysis of Pykacz above cited.

In order to better frame the question, we will report as the problem
was fronted by a few of these interpretations.

The Copenhagen interpretation says nothing about the cause of the
collapse. However, in its original formulations, wave functions aren't
regarded as ontologically real entities. Bohr did not think of quantum
measurement in terms of a collapse of the wave function\citep{QM-Copenhagen},
whereas Heisenberg considered wave functions representing a probability,
not an objective reality itself\citep{Heisenberg1958,QM-Copenhagen}.

According to the \emph{Von Neumann\textendash Wigner interpretation}\citep{Neumann1932,Wigner1967}
the collapse is caused by consciousness. Henry Stapp so summarizes
this perspective: \textquotedblleft In short, orthodox quantum mechanics
is Cartesian dualistic at the pragmatic/\emph{operational} level,
but mentalistic on the \emph{ontological} level\textquotedblright \citep{Stapp2009}.

The \emph{many-worlds interpretation}\citep{Everett1957,DeWitt1970}
asserts the objective reality of an universal wave function and denies
the actuality of wave function collapse: all possible alternative
histories and futures are real, each representing an actual ``world''
(or ``universe'').

In the \emph{De Broglie\textendash Bohm theory}\citep{Bohm1952,Bohm1952a},
in addition to a wave function on the space of all possible configurations,
it is also postulated an actual configuration that exists even when
unobserved. The time evolution of the configuration is defined by
the wave function via a guiding equation, while the wave function
evolves over time according the Schrödinger equation. According to
this interpretation the interaction with the environment during a
measurement procedure separates the wave packets in configuration
space, giving rise to an apparent wave function collapse. In such
interpretation wave functions are ontologically real.

The \emph{objective collapse theories} regard both the wave functions
and the process of collapse as ontologically objective. In such theories,
collapse occurs randomly (``spontaneous localization'') or when
some physical threshold is reached, giving the observer no special
role. The mechanism of collapse, that is not specified by standard
quantum mechanics, needs to be introduced \emph{ex novo}, if this
approach is correct. For this reason, Objective Collapse is more a
new theory than an interpretation. Examples include the \emph{Ghirardi-Rimini-Weber
theory}\citep{G.C.Ghirardi1985} and the \emph{Penrose interpretation}\citep{Penrose1989}.

Giuliani\textquoteright s position is also of interest, as he advocates
a form of ``\emph{tempered realism}''\citep{Giuliani2007}, an intermediate
position between pure \emph{instrumentalism} and \emph{classical realism}.
In this view, quantum theory should be restricted to the operational
description of state vectors and experimental outcomes, while the
question of whether physical quantities possess definite values prior
to measurement is regarded as an empirical, and in principle testable,
issue rather than a philosophical one.

\subsection{What it has been done in this paper}

The above picture is objectively discouraging and justify the widespread
attitude of physicists to simply ignore such issues. This is understandable
and it is the right thing to do when a conceptual node is beyond the
comprehension capabilities of the human of the time. When Isaac Newton
had to front the properties of gravity, in particular for what regards
distant action between material bodies, he was forced to a similar
withdrawn, stigmatized by his famous quote ``\emph{Hypotheses non
fingo}''\citep{Newton1726}. Notwithstanding this, the theory of
universal gravitation allowed the most accurate description of the
motion of celestial bodies for almost two centuries. In this long
period, the Newton doubt was completely forgot. But in 1915, Einstein
showed to the world, with his general theory of gravitation\citep{Einstein1915},
a much deeper insight of the Cosmos, giving an answer to the unsolvable
question of Newton!

In a similar manner, it is not inconceivable that also the measurement
problem and the plethora of interpretations to which it gave rise
is the symptom of more fundamental issues in the deeper structure
of the quantum mechanical theory. But what is the disease?

In my opinion, the problem sinks its roots in the heterogeneity of
the entities assumed as \emph{fundamental} and \emph{primitive} in
the formulation of Quantum Mechanics. As a matter of fact, they are
three and all of them ill-defined: quantum states (wave functions),
observables and observers. In particular, whereas, on the one hand,
it is recognized the fundamental role of of the observer in the definition
itself of the measurement outcomes of a physical quantities; on the
other, observers seem to have no special role in the determining the
expectation values of the quantities themselves. Besides, in order
to define the \textbf{one} expectation value of a physical quantity,
the theory requires \textbf{two} entities of different nature and
mathematical representation: a quantum state, often represented by
a wave function, and an observable, represented by an operator. This
leads to an ill definition of both the entities and to severe ambiguities
about their degrees of reality and objectivity. So in a group of interpretations
wave function are ontologically real and not in another; in a group
quantum state collapse is objective, in another is apparent and so
on\ldots{}

In this paper, after an introductory mathematical part, I will show
how it is possible to define quantum states, observables and expectation
values in term of an \textbf{only} mathematical entity: the algebra
of pseudo-observables, introduced in the first paper\citep{Piparo-I}
to justify the necessity of the Quantum Mechanics to describe the
measurable properties of a physical system. In doing this, it will
be also clearly stated the \emph{derived} and partly \emph{subjective}
character of the quantum states. In the conclusion, I will digress
on what we should understand by the term ``reality'' and to what
extent quantum states can be regarded as ``real''.

\section{Quantum states}

\subsection{Eigenvectors and eigenvalues\label{sec:Autovettori-ed-autovalori}}

After having indicated with $Z$ a generic element of the space of
the pseudo-observables $\mathbb{P}$, one is allowed to associate
to each pseudo-observable $P$ the two linear mappings:

\[
\varphi_{\mathrm{R}}\left(Z\right)\ideq P\,Z
\]
\[
\varphi_{\mathrm{L}}\left(Z\right)\ideq Z\,P
\]
in correspondence of which we can consider a \emph{right eigenvector}
and a \emph{right eigenvalue} relative to the pseudo-observable $P$,
i.e. a nonzero pseudo-observable $\Phi'$ and a complex number $\omega'$
such that it results:

\begin{equation}
P\,\Phi'=\omega'\,\Phi'\label{eq:Autovalore_Destro}
\end{equation}
and a \emph{left eigenvector} and a \emph{left eigenvalue} relative
to the pseudo-observable $P$, i.e. a nonzero pseudo-observable $\Phi''$
and a complex number $\omega''$ such that it results:

\begin{equation}
\Phi''\,P=\omega''\,\Phi''\text{ .}\label{eq:Autovalore_Sinistro}
\end{equation}
A pseudo-observable $\Psi$ that is both a right eigenvector, relative
to the eigenvalue $\omega'$, and left eigenvector, relative to the
eigenvalue $\omega''$, of the same pseudo-observable $P$ will be
called a \emph{bilateral eigenvector} relative to the pair of eigenvalues
$\left(\omega',\omega''\right)$. If $\Phi_{1}$ and $\Phi_{2}$ are
two bilateral eigenvectors of a pseudo-observable $P$ relative to
the same pair of eigenvalues $\left(\omega',\omega''\right)$, it
is immediately verified that every their linear combination, with
complex coefficients, is a bilateral eigenvector relative to the pair
of eigenvalues $\left(\omega',\omega''\right)$. The set of the bilateral
eigenvectors of a given pseudo-observable relative to the same pair
of eigenvalues $\left(\omega',\omega''\right)$ constitutes, therefore,
a vector subspace of $\mathbb{P}$, that we will call \emph{bilateral
eigenspace} relative to the pair of eigenvalues $\left(\omega',\omega''\right)$
and that we will denote with $\mathbb{E}\left(\omega',\omega''\right)$.

Note that the equations \eqref{eq:Autovalore_Destro} and \eqref{eq:Autovalore_Sinistro}
certainly admit nonzero solutions in the case in which $P$ is an
observable, since they are satisfied by the projectors of the basis
associated to $P$ and by its spectral coefficients, as shown by equation
(10) in the first paper, of which \eqref{eq:Autovalore_Destro}
and \eqref{eq:Autovalore_Sinistro} represent a generalization.

Let's, from now on, suppose that $P$ is equal to an observable $O$.

If $\left\{ I_{j}\right\} $ is a basis of pairwise orthogonal primitive
projectors associated to a complete set of observables compatible
one with each other and with $O$, the dyads associated with all possible
pairs of such projectors, as shown in subsection 4.2 in the
first paper, constitute a basis and are bilateral eigenvectors of
$O$. In fact, one has:

\begin{equation}
O\,\Gamma_{jk}=OI_{j}C_{jk}I_{k}=o_{j}\,I_{j}C_{jk}I_{k}=o_{j}\,\Gamma_{jk}\label{eq:Diade_Autov_dx}
\end{equation}
and
\begin{equation}
\Gamma_{jk}\,O=I_{j}C_{jk}I_{k}O=I_{j}C_{jk}\left(OI_{k}\right)=o_{k}\,I_{j}C_{jk}I_{k}=o_{k}\,\Gamma_{jk}\label{eq:Diade_Autov_sx}
\end{equation}
that show as the dyad $\Gamma_{jk}$ is a bilateral eigenvector of
$O$ relative to the pair of eigenvalues $\left(o_{j},o_{k}\right)$.

Let $\Phi$ be a right eigenvector of an observable $O$, relative
to the right eigenvalue $\omega$. By expressing $\Phi$ as a linear
combination of the dyads $\Gamma_{jk}$ associated to the pairs of
primitive projectors of a basis $\left\{ I_{j}\right\} $ compatibles
with $O$, making use of equation (46) in the first paper, one
has:

\begin{equation}
\Phi=\sum_{j,k}\varpi_{jk}\,\Gamma_{jk}\label{eq:Scomp_Psi}
\end{equation}
According to \eqref{eq:Diade_Autov_dx}, besides, it results:
\[
O\,\Phi=\sum_{j,k}\varpi_{jk}\,O\Gamma_{jk}=\sum_{j,k}o_{j}\varpi_{jk}\,\Gamma_{jk}
\]
which substituted together with \eqref{eq:Scomp_Psi} in \eqref{eq:Autovalore_Destro}
provides the relation:
\[
\sum_{j,k}\left(\omega-o_{j}\right)\varpi_{jk}\,\Gamma_{jk}=0
\]
that, for the linear independence of the dyads of the basis, is equivalent
to the condition:
\begin{equation}
\left(\omega-o_{j}\right)\varpi_{jk}=0\;\Rightarrow\;\varpi_{jk}\neq0\,\rightarrow\,\omega=o_{j}\text{ .}\label{eq:Equaz_Autovalori_dx}
\end{equation}
According to \eqref{eq:Equaz_Autovalori_dx}, the right eigenvalues
of $O$ are all and only the terms of its spectrum. Each right eigenvector
relative to a certain right eigenvalue $o$ of $O,$ besides, will
be given by a linear combination of the dyads that contain as first
factor the primitive projectors of a basis compatibles with $O$ whose
spectral coefficient in the decomposition (9) in the first paper
is equal to $o$. Similarly, for the left eigenvectors and eigenvalues.
It is, therefore, possible to affirm that, indicated with $o_{j}$
the spectral coefficient, in the $O$ decomposition (9) in the
first paper, of the primitive projector $I_{j}$ belonging to a basis
compatible with $O$, the bilateral eigenspace $\mathbb{E}\left(o',o''\right)$
admits as a basis the set of dyads: $\left\{ \Gamma_{jk}\,|\,o_{j}=o'\,\wedge\,o_{k}=o''\right\} $.
If the multiplicity of $o'$ is $m_{o'}$ and the multiplicity of
$o''$ is $m_{o''}$, the dimension of the bilateral eigenspace $\mathbb{E}\left(o',o''\right)$,
therefore, is:
\begin{equation}
\dim\left(\mathbb{E}\left(o',o''\right)\right)=m_{o'}\,m_{o''}\text{ .}\label{eq:Dim_Autosp}
\end{equation}
If, in particular, one considers: $o'=o''=o$, where $o$ is a whatever
outcome of $O$, one has:
\begin{equation}
\dim\left(\mathbb{E}\left(o,o\right)\right)=m_{o}^{2}\text{ .}\label{eq:Dim_Autosp_Diag}
\end{equation}
Since the dimension of an eigenspace is an intrinsic feature of the
observable $O$, independent from the choice of the basis $\left\{ I_{j}\right\} $
of primitive projectors, it is concluded that also \emph{the multiplicities
of the terms of the spectrum of $O$ are intrinsic characteristics
of the observable}.

\subsection{The trace of a pseudo-observable\label{sec:Traccia}}

According to what demonstrated in subsection 2.3 in the first
paper \citep{Piparo-I} and at the end of subsection \ref{sec:Autovettori-ed-autovalori},
both the terms of the spectrum of an observable $O$ and the relative
multiplicities are intrinsic characteristics of the observable itself.

We now define the \emph{trace} functional ($\tr$) of an observable
$O$ as the sum of all its spectral coefficients, $o_{j}$, with respect
to a basis $\left\{ I_{j}\right\} $ of pairwise orthogonal primitive
projectors compatible with it, i.e.:

\begin{equation}
\tr\left(O\right)\ideq\sum_{j}o_{j}\label{eq:Def_Traccia}
\end{equation}
The trace of an observable, that is evidently given also by the sum
of the products of the distinct terms of the observable spectrum by
their relative multiplicities, is, for what was exposed at the beginning
of this subsection, uniquely defined, since it can be expressed in
terms of intrinsic characteristics of the observable.

Consider now a dyad basis, $\left\{ \Gamma'_{ll'}\right\} $,whose
elements are generally not compatible with the projectors $I_{j}$,
and let be $\left\{ \Gamma_{jl}\right\} $ a dyad basis associated
to the primitive projector basis $\left\{ I_{j}\right\} $. Accord
to equation (59) in the first paper, one has

\begin{equation}
I_{j}=\Gamma_{jj}=\sum_{l,l'}\omega_{lj}\omega_{l'j}^{*}\,\Gamma'_{ll'}\label{eq:Scomp_Ind_Diadi}
\end{equation}
where the complex constants $\omega_{lj}$ are the components of the
pseudo-observable of the basis changing. By substitution of \eqref{eq:Scomp_Ind_Diadi}
in the decomposition of the observable $O$ according the basis $\left\{ I_{j}\right\} $
of primitive projectors, equation (9) in the first paper, one
obtains:
\begin{equation}
O=\sum_{l,l'}\left(\sum_{j}\omega_{lj}\omega_{l'j}^{*}\,o_{j}\right)\Gamma'_{ll'}=\sum_{l,l'}o'_{ll'}\,\Gamma'_{ll'}\label{eq:Scomp_Diadica_Oss}
\end{equation}
where the coefficient:
\[
o'_{ll'}\ideq\sum_{j}\omega_{lj}\omega_{l'j}^{*}\,o_{j}
\]
represents the component of $O$ relative to the dyad $\Gamma'_{ll'}$.
We will, now, demonstrate that the trace of an observable is given
by the sum of its diagonal components, i.e. it coincides with the
trace of the matrix associated with $O$. In fact, according to \eqref{eq:Scomp_Diadica_Oss}
and making use of equation (64) in the first paper, one has:
\[
\sum_{l}o'_{ll}=\sum_{l,j}\omega_{lj}\omega_{l'j}^{*}\,o_{j}=\sum_{j}o_{j}\sum_{l}\omega_{lj}^{*}\omega_{lj}=\sum_{j}o_{j}=\tr\left(O\right)\text{ .}
\]
In general, therefore, we can \emph{define the trace functional of
a pseudo-observable as the sum of its diagonal components with respect
to a whatever dyad basis}. The definition is well placed, since if
$P$ is a generic pseudo-observable, of real part $\rp P$ and imaginary
part $\imp P$, and $\left\{ \Gamma_{jk}\right\} $ is a whatever
dyad basis, called $\varpi_{jk}$, $\alpha_{jk}$ and $\beta_{jk}$
the component respectively of $P$, of its real part and of its imaginary
part relative to the dyad $\Gamma_{jk}$, one has:
\begin{equation}
\varpi_{jk}\ideq\alpha_{jk}+i\beta_{jk}\label{eq:Scomp_Comp_RI}
\end{equation}
by which it results:
\begin{equation}
\sum_{j}\varpi_{jj}=\sum_{j}\alpha_{jj}+i\sum_{j}\beta_{jj}=\tr\left(\rp P\right)+i\,\tr\left(\imp P\right)\text{ .}\label{eq:Traccia_RI}
\end{equation}
The so defined trace functional satisfies the three following properties:
\begin{enumerate}
\item For each pseudo-observable $P$ of real part $\rp P$ and imaginary
part $\imp P$, one has:
\[
\tr\left(P\right)\ideq\tr\left(\rp P\right)+i\,\tr\left(\imp P\right)\text{ .}
\]
\item For any two given pseudo-observables $P$ and $Q$, the trace of their
sum is equal to the sum of the their traces:
\[
\tr\left(P+Q\right)=\tr\left(P\right)+\tr\left(Q\right)\text{ .}
\]
\item If $P$ and $Q$ are two whatever pseudo-observables, one has:
\[
\tr\left(PQ\right)=\tr\left(QP\right)\text{ .}
\]
\end{enumerate}
The first property is an immediate consequence of the definition and
of \eqref{eq:Traccia_RI}. The second property follows immediately
by the definition making use of equation (48) in the first paper.
It allows to state that the trace is a \emph{linear functional}. For
what concerns the third property, indicated respectively by $\varpi_{jk}$
and $\theta_{jk}$ the components of $P$ and $Q$ relative to the
dyad $\Gamma_{jk}$ of a given basis, according to equation (49)
in the first paper, the component $\zeta_{jk}$ of the product $PQ$
relative to the dyad $\Gamma_{jk}$ is given by:
\begin{equation}
\zeta_{jk}=\sum_{l}\varpi_{jl}\,\theta_{lk}\label{eq:Comp_PQ}
\end{equation}
whereas the component $\xi_{jk}$ of the product $QP$ relative to
the dyad $\Gamma_{jk}$ is given by:
\begin{equation}
\xi_{jk}=\sum_{l}\theta_{jl}\,\varpi_{lk}\text{ .}\label{eq:Comp_QP}
\end{equation}
According to \eqref{eq:Comp_PQ} and \eqref{eq:Comp_QP}, therefore,
it results:
\[
\tr\left(PQ\right)=\sum_{j}\zeta_{jj}=\sum_{j,l}\varpi_{jl}\,\theta_{lj}=\sum_{j,l}\theta_{jl}\,\varpi_{lj}=\tr\left(QP\right)
\]
A first important consequence of these properties is that for every
pseudo-observable $Z$ it results:
\begin{equation}
\tr\left(Z^{\dagger}\right)=\tr\left(Z\right){}^{*}\text{ .}\label{eq:Traccia_Trasposta}
\end{equation}

Moreover, if $\gamma$ is a complex constant and $Z$ a pseudo-observable,
it is easily demonstrated that it results:
\begin{equation}
\tr\left(\gamma Z\right)=\gamma\,\tr\left(Z\right)\text{ .}\label{eq:Traccia_Prod_PC}
\end{equation}

An important consequence of the third property is that for each pair
of pseudo-observables $X$ and $Y$ it results:
\begin{equation}
\tr\left({\left[X,Y\right]}\right)=0\text{ .}\label{eq:Traccia_Commutatore}
\end{equation}

Note that if $O$ is a non-negative observable ($O\geq0$ ), then
the following important properties apply:
\begin{equation}
\tr\left(O\right)\geq0\label{eq:Traccia_Oss_Pos}
\end{equation}
and
\begin{equation}
\tr\left(O\right)=0\;\Leftrightarrow\;O=0\label{eq:Annull_Traccia_Oss_Pos}
\end{equation}
which is immediate to verify on the basis of the definition of the
trace of an observable.

Now consider the trace of a generic dyad $\Gamma_{jj'}$ of a basis
$\left\{ \Gamma_{jk}\right\} $. Since the only nonzero component
of $\Gamma_{jj'}$ according to the dyad basis is that relative to
itself, trivially equal to $1$, one has:
\begin{equation}
\tr\left(\Gamma_{jj'}\right)=\delta_{j,j'}\text{ .}\label{eq:Traccia_Diade}
\end{equation}
This relation allows one to characterize primitive projectors. Let,
in fact, $J$ be a projector. If $J$ is not primitive, it will be
expressible as a sum of $n$ pairwise orthogonal primitive projectors
$I_{j}$:
\begin{equation}
J=I_{1}+\cdots+I_{n}\text{ .}\label{eq:Scomp_Indic_Non_El}
\end{equation}
By definition, therefore, it will be $\tr\left(J\right)=n$. According
to this, the \emph{necessary and sufficient condition for a projector
$I$ to be primitive is that it results:}\textbf{\emph{ $\tr\left(I\right)=1$}}.

\subsection{Inner product of two pseudo-observables\label{sec:Prodotto-scalare-PO}}

The trace functional allows one to define a Frobenius-like \emph{inner
product} into the space $\mathbb{P}$ of the pseudo-observables. If
$X$ and $Y$ are two pseudo-observables, we define their inner product,
$\inprod XY$, as follows:
\begin{equation}
\inprod XY\ideq\tr\left(X^{\dagger}Y\right)\text{ .}\label{eq:Def_Prod_Scal}
\end{equation}
It's easy to verify that the so defined operation actually satisfies
the characteristic properties of a (complex) inner product, i.e. if
$\alpha$ and $\beta$ are two complex constants and $X$, $Y$ and
$Z$ three pseudo-observables, it holds that:
\begin{enumerate}
\item \emph{Anti-linearity} in the first argument:\\
$\inprod{\alpha\,X+\beta\,Z}Y=\alpha^{*}\inprod XY+\beta^{*}\inprod ZY$.
\item \emph{Linearity} in the second argument:\\
$\inprod X{\alpha\,Y+\beta\,Z}=\alpha\inprod XY+\beta\inprod XZ$.
\item \emph{Conjugate symmetry}:\\
$\inprod XY=\inprod YX^{*}$ .
\item \emph{Positive-definiteness}:\\
$\inprod XX\geq0$,\\
$\inprod XX=0$ iff $X=0$.
\end{enumerate}
The verification of the first three properties immediately follows
by the definition\eqref{eq:Def_Prod_Scal},while for the fourth one
must use the fifth and sixth property of the transposition and, in
the order, the properties \eqref{eq:Traccia_Oss_Pos} and \eqref{eq:Annull_Traccia_Oss_Pos}
of the trace.

The fourth property of the inner product allow one to introduce the
\emph{norm} $\left\Vert X\right\Vert $ of a generic pseudo-observable
$X$, defined as follows:
\begin{equation}
\left\Vert X\right\Vert \ideq\sqrt{\inprod XX}\label{eq:Def_Norma}
\end{equation}
that results always positive or zero, being equal to zero if and only
if $X=0$.

A normed space is also a \emph{metric space}, and we will assume the
further hypothesis that the space $\mathbb{P}$ of the pseudo-observables
is also \emph{complete}, that's to say that every \emph{Cauchy sequence}
of elements of $\mathbb{P}$ converges to an element of $\mathbb{P}$.
With respect to such a metric the space $\mathbb{P}$ of the pseudo-observables
is therefore an \emph{Hilbert space} formed by every formal expression
obtained by summing or multiplying observables, that we will suppose
to be, at most, a countable set, and by the limits of Cauchy sequences
of such expressions.

Further interesting properties of the the inner product, defined by
the \eqref{eq:Def_Prod_Scal}, are given by the following relations:

\begin{equation}
\inprod XY=\tr\left(X^{\dagger}Y\right)=\tr\left(\left(Y^{\dagger}\right)^{\dagger}X^{\dagger}\right)=\inprod{Y^{\dagger}}{X^{\dagger}}\text{ ,}\label{eq:Prod_Sc_Trasp}
\end{equation}
\begin{equation}
\inprod X{YZ}=\tr\left(X^{\dagger}YZ\right)=\tr\left(\left(Y^{\dagger}X\right)^{\dagger}Z\right)=\inprod{Y^{\dagger}X}Z\label{eq:Prod_Sc_Adj}
\end{equation}
that immediately follow by the trace properties.

Let now $O$ be a whatever observable and $\left\{ \Gamma_{jk}\right\} $
a dyad basis formed by bilateral eigenvectors of $O$. We want to
demonstrate that such dyads constitute an \emph{orthonormal basis}
for the space $\mathbb{P}$ of the pseudo-observables, that is that
for them it holds the following relation:

\begin{equation}
\inprod{\Gamma_{jk}}{\Gamma_{j'k'}}=\delta_{j,j'}\delta_{k,k'}\text{ .}\label{eq:Ortonorm_Diadi}
\end{equation}
In fact, one has:
\[
\inprod{\Gamma_{jk}}{\Gamma_{j'k'}}=\tr\left(\Gamma_{kj}\Gamma_{j'k'}\right)=\delta_{j,j'}\,\tr\left(\Gamma_{kj}\Gamma_{jk'}\right)=\delta_{j,j'}\,\tr\left(\Gamma_{kk'}\right)=\delta_{j,j'}\delta_{k,k'}
\]
where it was made use of the second and third properties of the dyads
(see paper \citep{Piparo-I}) and of the relation \eqref{eq:Traccia_Diade}.

We have so demonstrated that \emph{for each observable $O$ there
exists an orthonormal basis of the space}\textbf{\emph{ $\mathbb{P}$
}}\emph{of the pseudo-observables formed by (bilateral) eigenvectors
of $O$} (\textbf{\emph{first spectral theorem}}). Due to our assumptions
on the spectra of the observables, we will, therefore, assume that
the space $\mathbb{P}$ of the pseudo-observables is also \emph{separable},
i.e. it admits countable orthonormal bases.

The inner product introduced, finally, allows one to express the component
$\varpi_{jk}$ of a pseudo-observable $P$ relative the dyad $\Gamma_{jk}$
of a given basis in the form:

\begin{equation}
\varpi_{jk}=\inprod{\Gamma_{jk}}P\label{eq:Coeff_Scomp_PO}
\end{equation}
which is easily demonstrated starting from the decomposition (48)
in the first paper and making use of the orthonormality relations
\eqref{eq:Ortonorm_Diadi}.

\subsection{State vectors\label{sec:Vettori-di-Stato}}

Given a basis $\left\{ I_{j}\right\} $ of pairwise orthogonal primitive
projectors of a certain complete space of compatible observables,
it will be proved that there is a set $\left\{ \Psi_{j}\right\} $
of pseudo-observables such that the products $\Psi_{j}\Psi_{k}^{\dagger}$
form a dyad basis associated to the projector basis $\left\{ I_{j}\right\} $.
By indicating with $\left\{ \Gamma_{jk}\right\} $such a dyad basis,
this is equivalent to saying that, for each pair of indexes $j$ and
$k$, it must result:
\begin{equation}
\Psi_{j}\Psi_{k}^{\dagger}=\Gamma_{jk}\text{ .}\label{eq:Caratt_Vettori_Stato}
\end{equation}
The existence of the pseudo-observables satisfying the \eqref{eq:Caratt_Vettori_Stato}
is proved, as it is easy to verify, after arbitrarily choosing an
index $k_{0}$, by setting:
\[
\Psi_{j}\ideq\Gamma_{jk_{0}}\text{ .}
\]
The decomposition \eqref{eq:Caratt_Vettori_Stato} justifies the name
of dyad adopted in section 4 of the first paper.

For a given set of pseudo-observables satisfying the \eqref{eq:Caratt_Vettori_Stato},
we will call each its element a \emph{state vector}.

In the next subsection, it will be proved the fundamental \emph{theorem
of characterization of the sets of the state vectors}: \textbf{each
set of vector states is formed, for each state index $j$, by the
pseudo-observables of the form}:
\begin{equation}
\Psi_{j}=\Gamma_{jk_{0}}U\label{eq:Forma_Vettori_Stato}
\end{equation}
\textbf{where $k_{0}$ and $U$ are, respectively, an index and an
unitary pseudo-observable arbitrarily chosen}\footnote{\label{fn:Vettori-di-Stato}By fixing a whatever \emph{bra} $\bra{\upsilon}$,
normalized to one, the vector state $\Psi_{j}$, relative to state
of index $j$, thus corresponds to the operator $\ket j\bra{\upsilon}$
in the Dirac formulation Dirac, where also the \emph{ket} $\ket j$
is normalized to one. It is important to point out that is not possible
to individually define the state vectors, since the unitary pseudo-observable
$K$ which appears in their definition is the same for all of them.}. It is worth observing that the choice of the index $k_{0}$ in the
\eqref{eq:Forma_Vettori_Stato} is \textbf{irrelevant}. In fact, if
one regards a whatever different index $k_{1}$, one has:
\begin{equation}
\Psi_{j}=\Gamma_{jk_{0}}U=\Gamma_{jk_{1}}S_{k_{1}k_{0}}U=\Gamma_{jk_{1}}V\label{eq:Irrilev_K_0}
\end{equation}
being $S_{k_{1}k_{0}}$ the unitary pseudo-observable defined in equation
(65) of the first paper, and having put $V\ideq S_{k_{1}k_{0}}U$,
which is still an unitary pseudo-observable.

\subsection{Characterization of the state vectors\label{sec:Caratterizzazione-VS}}

We now prove the \emph{theorem of characterization of the sets of
the state vectors}. With the notations seen in the subsection \ref{sec:Vettori-di-Stato},
equation \eqref{eq:Caratt_Vettori_Stato}, and by making use of equation
(43) in the first paper, it is possible to express a state vector
$\Psi_{j}$ as a linear combination of the dyads of the basis $\left\{ \Gamma_{jk}\right\} $:
\begin{equation}
\Psi_{j}=\sum_{j',k}\psi_{j'k}\,\Gamma_{j'k}\label{eq:Scomp_Psi_j_1}
\end{equation}
where the coefficients $\psi_{j'k}$ are the components of $\Psi_{j}$
relative to the dyads of the basis. According to \eqref{eq:Caratt_Vettori_Stato},
considered for $k=j$, one has:
\begin{equation}
\Psi_{j}\Psi_{j}^{\dagger}=\Gamma_{jj}=I_{j}\text{ .}\label{eq:Scomp_Ind_VS}
\end{equation}
By substituting the \eqref{eq:Scomp_Psi_j_1} in the \eqref{eq:Scomp_Ind_VS}
and remembering the uniqueness of the decomposition of a pseudo-observable,
one has that in the \eqref{eq:Scomp_Psi_j_1} the only nonzero coefficients
are those for which it results $j'=j$, that is:
\begin{equation}
\Psi_{j}=\sum_{k}\psi_{jk}\,\Gamma_{jk}\text{ .}\label{eq:Scomp_Psi_j_2}
\end{equation}
According to the \eqref{eq:Scomp_Psi_j_2}, the \eqref{eq:Scomp_Ind_VS}
and to the properties of the dyads, besides, one has:
\begin{equation}
\Psi_{j}\Psi_{j}^{\dagger}\Psi_{j}=I_{j}\Psi_{j}=\Psi_{j}\text{ .}\label{eq:Prod_I_Psi_j}
\end{equation}
Consider, now, the following expression:
\begin{equation}
J_{j}\ideq\Psi_{j}^{\dagger}\Psi_{j}\text{ .}\label{eq:Def_J_j}
\end{equation}
According to the definition of state vector, to the \eqref{eq:Scomp_Ind_VS}
and to the \eqref{eq:Prod_I_Psi_j}, it results:
\[
J_{j}J_{j}^{\dagger}=\Psi_{j}^{\dagger}\left(\Psi_{j}\Psi_{j}^{\dagger}\right)\Psi_{j}=\Psi_{j}^{\dagger}I_{j}\Psi_{j}=\Psi_{j}^{\dagger}\Psi_{j}=J_{j}
\]
which implies, according to the equation (15) in the first paper,
that $J_{j}$ is a projector. Since, on the basis of the third property
of the trace and of the characterization of the primitive projectors,
given at the end of subsection \ref{sec:Traccia}, it besides results:
\[
\tr\left(J_{j}\right)=\tr\left(\Psi_{j}^{\dagger}\Psi_{j}\right)=\tr\left(\Psi_{j}\Psi_{j}^{\dagger}\right)=\tr\left(I_{j}\right)=1\text{ ,}
\]
 the projector $J_{j}$ is also primitive. Furthermore it is:
\begin{equation}
\Psi_{j}J_{j}=\Psi_{j}\Psi_{j}^{\dagger}\Psi_{j}=I_{j}\Psi_{j}=\Psi_{j}\label{eq:Prod_Psi_j_J}
\end{equation}
which follows immediately by the definition \eqref{eq:Def_J_j} and
by the \eqref{eq:Prod_I_Psi_j}.

Consider now a complete set of compatible observables containing $J_{j}$.
This set identifies a certain basis $\left\{ K_{k}\right\} $ of primitive
projectors, containing, on the basis of the procedure of construction
seen in subsection 2.7 of the first paper, $J_{j}$. By arbitrarily
choosing an index $k_{0},$ with a suitable reordering of the basis
elements, you can make sure that $J_{j}$ is the basis element of
index $k_{0}$:
\begin{equation}
K_{k_{0}}\ideq J_{j}\text{ .}\label{eq:Base_Indic_J_j}
\end{equation}
By indicating with $\left\{ \Lambda_{jk}\right\} $ the dyad basis
associated to $\left\{ K_{k}\right\} $, let $\Omega_{j}$ be the
pseudo-observable of the change from the basis $\left\{ \Lambda_{jk}\right\} $
to the basis $\left\{ \Gamma_{jk}\right\} $. So it results:
\begin{equation}
J_{j}=K_{k_{0}}=\Omega_{j}I_{k_{0}}\Omega_{j}^{\dagger}\text{ .}\label{eq:Camb_Base_J_I}
\end{equation}
According to such relation, to the \eqref{eq:Prod_I_Psi_j} and to
the \eqref{eq:Prod_Psi_j_J}, therefore one has:
\begin{equation}
\Psi_{j}=I_{j}\Psi_{j}J_{j}=\left(I_{j}\Psi_{j}\Omega_{j}I_{k_{0}}\right)\Omega_{j}^{\dagger}=\psi_{j}\Gamma_{jk_{0}}\Omega_{j}^{\dagger}\label{eq:Forma_Psi_j}
\end{equation}
having substituted the expression among brackets according to equation
(44) in the first paper, where $\psi_{j}$ is a suitable complex
coefficient (the dyadic component). By imposing the validity of the
\eqref{eq:Scomp_Ind_VS}, besides one finds:
\begin{equation}
\left|\psi_{j}\right|^{2}=1\text{ .}\label{eq:Norm_psi_tilde}
\end{equation}

Chosen a whatever index $k$, consider, now, the following expression:
\[
J_{j}J_{k}J_{j}=\Psi_{j}^{\dagger}\Psi_{j}\Psi_{k}^{\dagger}\Psi_{k}\Psi_{j}^{\dagger}\Psi_{j}=\Psi_{j}^{\dagger}\Gamma_{jk}\Gamma_{kj}\Psi_{j}=\Psi_{j}^{\dagger}I_{j}\Psi_{j}=\Psi_{j}^{\dagger}\Psi_{j}=J_{j}
\]
having made use of the \eqref{eq:Caratt_Vettori_Stato}, of the third
property of the dyads and of the \eqref{eq:Prod_I_Psi_j}. According
to the lemma proved in \ref{sec:Appendice}, for each pair of indexes
$j$ and $k$, $J_{j}=J_{k}$, i.e. the projectors $J_{j}$ are all
equals one each other. We can therefore assume that also the bases
of which they are part coincide one with each other, so that the pseudo-observable
of the change of the dyad basis too is the same for all of the values
of the state index $j$:
\begin{equation}
\Omega_{j}=\Omega\text{ .}\label{eq:Camb_Base_Cost}
\end{equation}
By substitution of such relation into the \eqref{eq:Forma_Psi_j}
and imposing the validity of the \eqref{eq:Caratt_Vettori_Stato}
for each pair of indexes $j$ and $k$, according to the \eqref{eq:Norm_psi_tilde},
one finds also that the dyadic components $\psi_{j}$ for all the
states are equal to a constant phase factor:
\begin{equation}
\psi_{j}=e^{i\vartheta}\text{ .}\label{eq:Fatt_Fase_Cost}
\end{equation}
By substituting the \eqref{eq:Camb_Base_Cost} and the \eqref{eq:Fatt_Fase_Cost}
into the \eqref{eq:Forma_Psi_j}, one lastly finds:
\[
\Psi_{j}=\Gamma_{jk_{0}}\left(e^{i\vartheta}\Omega^{\dagger}\right)
\]
That coincides with the \eqref{eq:Forma_Vettori_Stato} as soon as
you put $U\ideq e^{i\vartheta}\Omega^{\dagger}$, that is immediately
verified to be an unitary pseudo-observable.

State vectors verifying the \eqref{eq:Caratt_Vettori_Stato} for equivalent
dyad bases, belonging to different sets, corresponds one with each
other according an equivalence relation, as it is easy to verify,
and are therefore called \emph{equivalent}.

If $\Psi_{j}$ and $\Phi_{j}$ are two equivalent state vectors, according
to what has been demonstrated and to the (55) in the first paper,
it will be: $\Psi_{j}=\Gamma_{jk_{0}}U$ and $\Phi_{j}=e^{i\vartheta_{j}}\Gamma_{jk_{0}}V$,
where $U$ and $V$ are two unitary pseudo-observables and $\vartheta_{j}$
are arbitrary phase factors. So, by the properties of the unitary
pseudo-observables, one obtains:
\begin{equation}
\Phi_{j}=e^{i\vartheta_{j}}\Gamma_{jk_{0}}V=e^{i\vartheta_{j}}\Gamma_{jk_{0}}U\,U^{\dagger}V=e^{i\vartheta_{j}}\Psi_{j}\left(U^{\dagger}V\right)=e^{i\vartheta_{j}}\Psi_{j}Y\label{eq:Equiv_Vettori_Stato}
\end{equation}
being $Y\ideq U^{\dagger}V$ a suitable unitary pseudo-observable.
One can easily realize, by making use of the properties of the unitary
pseudo-observables, that by multiplying the elements of a set of state
vectors for arbitrary phase factors and, to the left, for a same unitary
pseudo-observable, it is obtained a set of state vectors equivalent
to the previous ones.

\subsection{The eigenstate space\label{sec:Spazio-autostati}}

Let's now look at some other important properties of the state vectors.
\begin{enumerate}
\item The state vectors are right eigenvectors for the observables of the
complete space of compatible observables having the basis $\left\{ I_{j}\right\} $
of primitive projectors. If, in fact, $O$ is an observable of this
space, according to \eqref{eq:Forma_Vettori_Stato}, one has:
\begin{equation}
O\,\Psi_{j}=O\,\Gamma_{jk_{0}}U=o_{j}\,\Gamma_{jk_{0}}U=o_{j}\,\Psi_{j}\text{ .}\label{eq:Autostati}
\end{equation}
For this reason, these pseudo-observables will be called also \emph{eigenstates}
of the observable $O$.
\item The state vectors form an \emph{orthonormal set}. In fact, it results:
\begin{equation}
\inprod{\Psi_{j}}{\Psi_{j'}}=\tr\left(\Psi_{j}^{\dagger}\Psi_{j'}\right)=\tr\left(\Psi_{j'}\Psi_{j}^{\dagger}\right)=\tr\left(\Gamma_{j'j}\right)=\delta_{j'j}\label{eq:Ortonorm_VS}
\end{equation}
as follows from the \eqref{eq:Caratt_Vettori_Stato} and the \eqref{eq:Traccia_Diade}.
\item By multiplying a state vector to the left for a whatever pseudo-observable
$P$, one obtains a linear combination of state vectors belonging
to the same set. Indicated with $\varpi_{j'k'}$ the dyadic component
of $P$ relative to the generic dyad $\Gamma_{j'k'}$, according to
the \eqref{eq:Forma_Vettori_Stato}, in fact, one has:
\begin{equation}
P\,\Psi_{j}=\sum_{j',k'}\varpi_{j'k'}\,\Gamma_{j'k'}\Gamma_{jk_{0}}U=\sum_{j'}\varpi_{j'j}\,\Gamma_{j'k_{0}}U=\sum_{j'}\varpi_{j'j}\,\Psi_{j'}\label{eq:Prod_PO_VS}
\end{equation}
having exploited the third property of the dyads.
\end{enumerate}
According to the first of the above demonstrated properties, the state
vectors of a given set are also \emph{simultaneous eigenstates} of
every set of observable of the complete space of compatible observables
spanned by the basis $\left\{ I_{j}\right\} $ of pairwise orthogonal
primitive projectors. In particular, they are simultaneous eigenstates
of a complete set of observables of the space. If we put these observables
in a $n$-tuple $\mathbf{O}$, after having indicated with $\mathbf{o}_{j}$
the $n$-tuple of eigenvalues of the observables of $\mathbf{O}$
relative to the state index $j$, it is:
\begin{equation}
\mathbf{O}\,\Psi_{j}=\mathbf{o}_{j}\,\Psi_{j}\label{eq:Eq_Autostati_Vettoriale}
\end{equation}
where the product of a $n$-tuple of observables by an observable,
in the left-hand side, is meant as the $n$-tuple having as components
the products of the various observables composing the $n$-tuple itself
by the observable by which the $n$-tuple is multiplied. Note that
to each $n$-tuple $\mathbf{o}_{j}$ of eigenvalues corresponds, apart
from an equivalence i.e. of the product by a phase factor, an only
eigenstate $\Psi_{j}$ of the set and vice versa, so that the correspondence
is \emph{biunivocal}, apart from an equivalence. On the other side,
if one considers a set $\left\{ \Psi_{j}\right\} $ of state vectors,
this can be always thought as formed by simultaneous eigenstates of
a complete set of the space spanned by the basis $\left\{ I_{j}\right\} $
of primitive projectors and therefore each state vector of he set
will be in a biunivocal correspondence, apart from an equivalence,
with the $n$-tuple of the observables of the complete set of the
space spanned by $\left\{ I_{j}\right\} $. For this reason, from
now on, we will call $\left\{ \Psi_{j}\right\} $ an \emph{eigenstate
set}.

It is worth observing that the first property is generalizable to
the case of a whatever complex function  $\phi\left(\mathbf{O}\right)$
of the $n$-tuple $\mathbf{O}$ of a complete set of observable. In
such a case, however, the eigenvalues $\phi\left(\mathbf{o}_{j}\right)$
will be complex.

According to the properties of the state vectors, we can therefore
state that the space $\mathbb{E}\ideq\sspan[\left(\left\{ \Psi_{j}\right\} \right)]$
of the linear combinations of the state vectors of a given set is
\emph{closed} under the left multiplication by a pseudo-observable
of the space $\mathbb{P}$ and admits as an orthonormal basis  the
eigenstate set $\left\{ \Psi_{j}\right\} $. The space $\mathbb{E}$
will be so called the \emph{eigenstate space}.

We now demonstrate the important \emph{eigenstate space characterization
theorem}: \textbf{each eigenstate set is equivalent to an orthonormal
basis of $\mathfrak{\mathbb{E}}$; each orthonormal basis of $\mathfrak{\mathbb{E}}$
is an eigenstate set}.

In order to prove the first part of the theorem, let $\left\{ \Phi_{j}\right\} $
be an eigenstate set and $\left\{ \Lambda_{jk}\right\} $ the corresponding
dyad basis. By indicating with $\Omega$ the unitary pseudo-observable
of the change from the basis $\left\{ \Lambda_{jk}\right\} $ to the
basis $\left\{ \Gamma_{jk}\right\} $, according to the equation (64)
in the first paper and to the \eqref{eq:Caratt_Vettori_Stato}, one
has:
\[
\Lambda_{jk}=\Omega\Gamma_{jk}\Omega^{\dagger}=\Omega\Psi_{j}\Psi_{k}^{\dagger}\Omega^{\dagger}=\left(\Omega\Psi_{j}\right)\left(\Omega\Psi_{k}\right)^{\dagger}=\Phi_{j}\Phi_{k}^{\dagger}
\]
where it was put:
\begin{equation}
\Phi_{j}\ideq\Omega\Psi_{j}\label{eq:Camb_Autostati}
\end{equation}
which is an element of $\sspan[\left(\left\{ \Psi_{k}\right\} \right)]$
by virtue of the third property of the state vectors. The set $\left\{ \Phi_{j}\right\} $
of pseudo-observables is then, by definition, equation \eqref{eq:Caratt_Vettori_Stato},
an eigenstate set corresponding to the dyad basis $\left\{ \Lambda_{jk}\right\} $
and therefore equivalent to every other eigenstate set relative to
the same dyad basis. Since $\Omega$ is invertible, as unitary, the
eigenstates $\Phi_{j}$ span the space $\mathfrak{\mathbb{E}}$ and
since they form an orthonormal set, according to the second property
of the state vectors, and are therefore linearly independent, they
form an orthonormal basis of $\mathfrak{\mathbb{E}}$. It is worth
noting that the equation \eqref{eq:Camb_Autostati} may, therefore,
be interpreted as expressing the change from the eigenstate basis
$\left\{ \Phi_{j}\right\} $ to the eigenstate basis $\left\{ \Psi_{j}\right\} $.

For what concerns the demonstration of the second part of the theorem,
we firstly remember the following important property: if $\Phi$ is
an element of the eigenstate space $\mathfrak{\mathbb{E}}=\sspan[\left(\left\{ \Psi_{j}\right\} \right)]$,
then it results:
\begin{equation}
\Phi=\sum_{j}\inprod{\Psi_{j}}{\Phi}\Psi_{j}\text{ .}\label{eq:Sviluppo_Autostati}
\end{equation}

This being premised, let $\left\{ \Phi_{j}\right\} $ an orthonormal
basis of $\mathbb{E}$. According to the \eqref{eq:Sviluppo_Autostati},
for each index $j$, one has:
\begin{equation}
\Phi_{j}=\sum_{k}\varphi_{kj}\,\Psi_{k}\quad\mbox{with}\quad\varphi_{kj}\ideq\inprod{\Psi_{k}}{\Phi_{j}}\text{ .}\label{eq:Sviluppo_Phi_j}
\end{equation}
But it is also:
\begin{equation}
\Psi_{k}=\sum_{l}\inprod{\Phi_{l}}{\Psi_{k}}\Phi_{l}=\sum_{l}\varphi_{kl}^{*}\,\Phi_{l}\label{eq:Sviluppo_Psi_k}
\end{equation}
where it was made use of the third property of the inner product.
By substitution of the \eqref{eq:Sviluppo_Psi_k} into the \eqref{eq:Sviluppo_Phi_j}
and vice versa and exploiting the linearly independence of the elements
of the basis, one obtains:
\begin{equation}
\sum_{k}\varphi_{kj}^{*}\varphi_{kl}=\sum_{k}\varphi_{jk}\varphi_{lk}^{*}=\delta_{j,l}\label{eq:Relazione_unitaria}
\end{equation}
where, as usual, $\delta_{j,l}$ is the Kronecker symbol. Observe,
now, that, according to the \eqref{eq:Sviluppo_Phi_j} and the \eqref{eq:Caratt_Vettori_Stato},
it results:
\begin{equation}
\Phi_{j}\Phi_{j'}^{\dagger}=\sum_{k,l}\varphi_{kj}\varphi_{lj'}^{*}\Gamma_{kl}=\Omega\Gamma_{jj'}\Omega^{\dagger}\label{eq:Prod_Phi_j_Phi_j1}
\end{equation}
having put:
\begin{equation}
\Omega\ideq\sum_{k,k'}\varphi_{kk'}\,\Gamma_{kk'}\label{eq:Def_Omega}
\end{equation}
which, by virtue of the \eqref{eq:Relazione_unitaria}, results to
be unitary. The \eqref{eq:Prod_Phi_j_Phi_j1} may therefore interpreted
as expressing the change from a certain dyad basis $\left\{ \Lambda_{jj'}\right\} $
to the dyad basis $\left\{ \Gamma_{jj'}\right\} $, with:
\[
\Phi_{j}\Phi_{j'}^{\dagger}=\Lambda_{jj'}\text{ .}
\]
By definition, the elements of the orthonormal basis $\left\{ \Phi_{j}\right\} $
so form an eigenstate set, that is \textit{quod erat demonstrandum}.

An important consequence of what above demonstrated is the \emph{state
superposition theorem}:\textbf{ each nonzero element $\Phi$, with
finite norm, of the space $\mathbb{E}$ is proportional to an eigenstate}.
This theorem is equivalent to the fundamental \emph{superposition
principle} in the Dirac formulation of quantum mechanics\citep{Dirac1930}.
It is, in fact, always possible, for instance by following the well-known
\emph{Gram-Schmidt process}, to build an orthonormal basis of $\mathbb{E}$.
The assert follows, therefore, as an immediate consequence of the
second part of the eigenstate space characterization theorem.

According to what above demonstrated, one may conclude that \textbf{the
eigenstate space $\mathfrak{\mathbb{E}}$ is spanned by an orthonormal
set of right eigenvectors of every observable $O$ (}\textbf{\emph{second
spectral theorem}}\textbf{)}.

In any case, as it is well known, the inner product of two right eigenvectors
of an observable relative to different eigenvalues is always equal
to zero (the eigenvectors are \emph{orthogonal} to each other). In
fact, if $\Phi_{1}$ and $\Phi_{2}$ are two right eigenvectors of
an observable $O$ respectively relative to the eigenvalues $o_{1}$
and $o_{2}$, one has $O\Phi_{1}=o_{1}\,\Phi_{1}$ and $O\Phi_{2}=o_{2}\,\Phi_{2}$.
According to the hypotheses assumed and to the properties of the inner
product, it is: $\inprod{\Phi_{1}}{O\Phi_{2}}=o_{2}\left\langle \Phi_{1},\Phi_{2}\right\rangle $,
whereas, by virtue of the  \eqref{eq:Prod_Sc_Adj}, it results: $\inprod{\Phi_{1}}{O\Phi_{2}}=\inprod{O\Phi_{1}}{\Phi_{2}}=o_{1}\left\langle \Phi_{1},\Phi_{2}\right\rangle $,
that, being by hypothesis $o_{1}\neq o_{2}$, imply: $\left\langle \Phi_{1},\Phi_{2}\right\rangle =0$.

The state vectors introduced in this section corresponds, apart from
an equivalence relation, to the analogue ones in the Dirac formulation
of quantum mechanics. A remarkable difference is, however, that while
in the Dirac formulation of quantum mechanics the state vectors are
introduced axiomatically and not without problems for what regards
the physical interpretation, especially for what concern the superposition
principle, in our framework the eigenstates are merely auxiliary pseudo-observables
that synthesize the statistical properties of the physical system
with respect to a given complete set of compatible observables and
that allow to simplify the equations describing the relationships
and the transformation laws of the observables. Particular attention
should also be paid to the circumstance, already highlighted in the
note \vref{fn:Vettori-di-Stato}, that the eigenstates cannot be defined
individually, but only as a whole relative to all the possible states
of the physical system. In this context, the simplifications in the
formalism introduced through the eigenstates are paid in terms of
the \emph{interpretability} of the results, being the link between
eigenstates and observation rather indirect.

It is finally worth observing that, due to equation \eqref{eq:Caratt_Vettori_Stato},
the state vectors forms a \textbf{set of }\textbf{\emph{generators}}
for the whole C{*}-algebra $\mathbb{P}$ of the pseudo-observables. 

\section{Expectation values and wave functions}

\subsection{Expectation value of an observable\label{sec:Valori-medi}}

By performing repeated measurements of an observable relative to a
given system in a certain state, you will usually obtain a set of
different outcomes. Therefore, it is necessary to sum up such measures
into a single estimate: the \emph{expectation value} $\left\langle O\right\rangle $
of the observable $O$. The result of the measurement process of a
observable is precisely given by this estimate.

The expectation value must satisfy the following fundamental properties:
\begin{enumerate}
\item For two given observables $O_{1}$ and $O_{2}$, the expectation value
of their sum is equal to the sum of their expectation values:\\
$\left\langle O_{1}+O_{2}\right\rangle =\left\langle O_{1}\right\rangle +\left\langle O_{2}\right\rangle $.
\item Given an observable $O$ and a real constant $c$, the expectation
value of the product of the constant by the observable is equal to
the product of the constant by the expectation value of the observable:\\
$\left\langle c\,O\right\rangle =c\left\langle O\right\rangle $.
\item For two given comparable observables $O_{1}$ and $O_{2}$, with $O_{1}\leq O_{2}$,
the expectation values follow the same order of the two observables,
so one has:\\
$\left\langle O_{1}\right\rangle \leq\left\langle O_{2}\right\rangle $.
\item If $c$ is a constant observable, its expectation value coincides
with the value of the constant itself:\\
$\left\langle c\right\rangle =c$.
\end{enumerate}
The first two properties express the \emph{linearity} of the expectation
value.

If one considers an observable $O$, belonging to a complete space
of compatible observables that admits the basis $\left\{ I_{j}\right\} $
of primitive projectors, and decomposes it according such a basis
in the form of equation (9) of the first paper, with the spectral
coefficients $o_{j}$, the linearity of expectation value implies
that it results:

\begin{equation}
\left\langle O\right\rangle =\sum_{j}o_{j}p_{j}\label{eq:Scomp_Val_Med}
\end{equation}
where:
\begin{equation}
p_{j}\ideq\left\langle I_{j}\right\rangle \text{ .}\label{eq:Def_Prob}
\end{equation}

\noindent By virtue of the third and the fourth properties of the
expectation value and remembering the relation (1) seen in the
first paper, one also has:
\begin{equation}
0\leq p_{j}\leq1\text{ .}\label{eq:Lim_Prob}
\end{equation}
The linearity of expectation value and the closure relation (7)
in the first paper, besides, imply that it results:
\begin{equation}
\sum_{j}p_{j}=1\text{ .}\label{eq:Chiusura_Prob}
\end{equation}
In force of the relations \eqref{eq:Lim_Prob} and \eqref{eq:Chiusura_Prob},
each coefficient $p_{j}$ therefore represents the \emph{probability}
of the elementary event associated to the projector $I_{j}$.

The \emph{probability distribution} $\left\{ p_{j}\right\} $ depends
on the system state and on the complete set of compatible observables
one is considering. \textbf{But, more subtly, it depends also on a
personal choice of the observer}. In fact, as de Finetti stated: ``PROBABILITY
DOES NOT EXIST'', in the sense of ``objective'' probability, ``only
\emph{subjective} probabilities exist \textendash{} that is, the \emph{degree
of belief} in the occurrence of an event attributed by a given person
at a given instant and with a given set of information''\citep{DeFinetti1974}.

Chosen a certain complete set of compatible observables, the state
of the system is therefore perfectly identified only after the \textbf{assignment}
of a such probability distribution. In this sense, we can agree also
with Fuchs' quote: ``QUANTUM STATES DO NOT EXIST''\citep{Fuchs2010}.

The states for which all but one of the probabilities are zero, that
is for which it holds: $p_{j}=\delta_{j,j'}$, are \emph{pure states.}
The states for which this instead is not true, are \emph{mixed states}.
It should be pointed out that whereas a mixed state describes the
situation of a physical system relative to the possible measurement
of a observer belonging to a given complete space of compatible observables,
a pure state describes the situation of a physical system after the
measurement has occurred and is, therefore, determined both by the
system and by the measurement process.

We will now demonstrate the following important property: \emph{if
the expectation value of an observable is zero for every state of
a physical system, then the observable is equal to $0$}. Let $O$
be an observable whose expectation value is zero for every state of
the physical system. As a consequence of the \eqref{eq:Scomp_Val_Med},
it will therefore be:

\noindent
\begin{equation}
\sum_{j}o_{j}p_{j}=0\label{eq:Media_Nulla}
\end{equation}
for any possible choice of the probabilities $p_{j}$, under the constrains
given by \eqref{eq:Lim_Prob} and by \eqref{eq:Chiusura_Prob}. Fixed
a whatever index $j'$, let's consider the pure state for which it
results: $p_{j}=\delta_{j,j'}$. From \eqref{eq:Media_Nulla}, it
therefore follows: $o_{j'}=0$. By the arbitrariness of the index
$j'$, it follows that all possible outcomes of $O$ are equal to
zero and therefore, for the equivalence criterion between observables,
that $O$ is the null observable.

As a corollary of this property, one has that \emph{if the expectation
values of two observables coincide for every state of a physical system,
then the two observables are equivalent}. In order to demonstrate
this statement, it suffices to consider the difference between the
two observable and to note that, by virtue of the linearity of the
expectation value and of the property just demonstrated, it have necessarily
to be the null observable.

It is possible to characterize the state of a physical system, relative
to the observation of a given complete space of compatible observables
having the basis $\left\{ I_{j}\right\} $ of pairwise orthogonal
primitive indicators, with obvious reference to the \emph{density
matrix}\citep{Neumann1927} and the \emph{density operator}\citep{Fano1957},
through the \textbf{\emph{density observable}}:
\begin{equation}
D\ideq\sum_{j}p_{j}\,I_{j}\label{eq:Def_Dens}
\end{equation}
where the probabilities $p_{j}$ are given by \eqref{eq:Def_Prob}.
The \emph{density observable} is the observable whose spectrum is
formed by the probabilities associated to the projectors of the basis.
Let now $O$ be a generic observable of the considered space. By making
use of the equations: (9), (2) and (6) of the first
paper, one obtains the following identity:
\begin{equation}
OD=\sum_{j}o_{j}p_{j}\,I_{j}\text{ .}\label{eq:Prod_Oss_Dens}
\end{equation}

Exploiting the definition of the trace functional \eqref{eq:Def_Traccia}
and the equations \eqref{eq:Prod_Oss_Dens} and\eqref{eq:Scomp_Val_Med},
one has, for whatever observable $O$ belonging to a given complete
space of compatible observables:
\begin{equation}
\tr\left(OD\right)=\sum_{j}o_{j}p_{j}=\left\langle O\right\rangle \text{ .}\label{eq:Traccia_Val_Med}
\end{equation}
Equation \eqref{eq:Traccia_Val_Med} shows that in order to calculate
the expectation values it is actually necessary the only knowledge
of the density observable.

\subsection{Expectation values of pseudo-observables\label{sec:Valori-medi-PO}}

We will now extend the concept of expectation value to the pseudo-observables
too.

Let's suppose having characterized the state of a physical system
through a \emph{maximum observation}, i.e. the measurement of a complete
set of compatible observables. These observables identify a certain
complete space $\mathbb{A}$ of compatible observables with a basis
$\left\{ I_{\mathbb{A},j}\right\} $ of pairwise orthogonal primitive
indicators. As outlined in subsection \ref{sec:Valori-medi}, such
a state is characterized by of the density observable:
\begin{equation}
D_{\mathbb{A}}=\sum_{j}p_{\mathbb{A},j}\,I_{\mathbb{A},j}\label{eq:Dens_A}
\end{equation}
where the probabilities $p_{\mathbb{A},j}$ are equal to the expectation
values of the basis projectors:
\begin{equation}
p_{\mathbb{A},j}=\left\langle I_{\mathbb{A},j}\right\rangle \text{ .}\label{eq:Probab_A}
\end{equation}

If now one measures an observable $B$ incompatible with at least
one of the observables of $\mathbb{A}$, the information obtained
in previous measurements must affect the expectation value of $B,$
that must, in some ways, be brought back to the complete set of information
acquired in previous measurements. The expectation value of $B$ will
be, therefore, given by the expectation value of an its suitable «projection»
on the space $\mathbb{A}$. To this end, after indicating with $B_{\mathbb{A},j}$
the projection of the observable $B$ according the projector $I_{\mathbb{A},j}$
(see subsection 4.1 in the first paper), given by:
\begin{equation}
B_{\mathbb{A},j}\ideq I_{\mathbb{A},j}BI_{\mathbb{A},j}=b_{\mathbb{A},j}\,I_{\mathbb{A},j}\text{ ,}\label{eq:Proiezione_Aj}
\end{equation}
where $b_{\mathbb{A},j}$ is a suitable constant, it is natural to
define as the \emph{projection}\textbf{\textit{\emph{ }}}\emph{$B_{\mathbb{A}}$
of an observable $B$ on a complete space $\mathbb{A}$ of compatible
observables}, the observable given by:
\begin{equation}
B_{\mathbb{A}}\ideq\sum_{j}B_{\mathbb{A},j}\text{ .}\label{eq:Def_Proiezione}
\end{equation}
The projection is an observable compatible with those of the space
$\mathbb{A}$, whose expectation value gives those of the observable
$B$ \emph{conditioned} by the previous maximum observation relative
to the space $\mathbb{A}$ . By indicating with $\left\langle B\right\rangle _{\mathbb{A}}$
such a \emph{conditioned expectation value} and remembering the \eqref{eq:Traccia_Val_Med},
on the basis of \eqref{eq:Dens_A}, \eqref{eq:Probab_A}, \eqref{eq:Proiezione_Aj}
and \eqref{eq:Def_Proiezione}, therefore we will set:
\begin{equation}
\left\langle B\right\rangle _{\mathbb{A}}\ideq\left\langle B_{\mathbb{A}}\right\rangle =\sum_{j}b_{\mathbb{A},j}p_{j}=\tr\left(D_{\mathbb{A}}B_{\mathbb{A}}\right)\text{ .}\label{eq:Valore_Medio_Cond_1}
\end{equation}

The calculation of the expectation value of $B$ can, however, be
led also in the following way. After having indicated with $\mathbb{B}$
a complete space of compatible observables, spanned by the basis $\left\{ I_{\mathbb{B},k}\right\} $
of pairwise orthogonal primitive projectors $\left\{ I_{\mathbb{B},k}\right\} $,
generated from the observable $B$ and all the observables of $\mathbb{A}$
compatible with $B$, let's consider the projection $D_{\mathbb{A},\mathbb{B}}$
of the density observable $D_{\mathbb{A}}$ on the space $\mathbb{B}$:
\begin{equation}
D_{\mathbb{A},\mathbb{B}}=\sum_{k}I_{\mathbb{B},k}D_{\mathbb{A}}I_{\mathbb{B},k}=\sum_{j,k}p_{\mathbb{A},j}\,I_{\mathbb{B},k}I_{\mathbb{A},j}I_{\mathbb{B},k}\label{eq:Proiezione_Dens}
\end{equation}
whose spectrum, for what seen above, is formed by the probabilities
of getting the events associated with the projectors $I_{\mathbb{B},k}$
conditioned by a previous maximum observation relative to the space
$\mathbb{A}$. The observable $D_{\mathbb{A},\mathbb{B}}$ represents
the new density describing the statistical properties of the physical
system \textbf{after} the measuring of $B$.

In order to prove this, we observe that, according the properties
of the projection of an observable, one must have:
\[
I_{\mathbb{B},k}I_{\mathbb{A},j}I_{\mathbb{B},k}=p_{\mathbb{B}j,k}\,I_{\mathbb{B},k}
\]
where, again, $p_{\mathbb{B}j,k}$ is a suitable constant dependent
on both indexes $j$ and $k$. By making use of the definition and
the properties of the trace functional, of the idempotency relation
(2) in the first paper and of the positive-definiteness of the
inner product, one has:
\[
p_{\mathbb{B}j,k}=\tr\left(I_{\mathbb{B},k}I_{\mathbb{A},j}I_{\mathbb{B},k}\right)=\inprod{I_{\mathbb{A},j}I_{\mathbb{B},k}}{I_{\mathbb{A},j}I_{\mathbb{B},k}}\geq0\text{ .}
\]
By virtue of the well-known \emph{Cauchy-Schwarz inequality}, it also
results:
\begin{eqnarray*}
p_{\mathbb{B}j,k} & = & \tr\left(I_{\mathbb{B},k}I_{\mathbb{A},j}I_{\mathbb{B},k}\right)=\tr\left(I_{\mathbb{B},k}\left(I_{\mathbb{A},j}I_{\mathbb{B},k}\right)\right)=\tr\left(\left(I_{\mathbb{A},j}I_{\mathbb{B},k}\right)I_{\mathbb{B},k}\right)=\\
 & = & \tr\left(I_{\mathbb{A},j}I_{\mathbb{B},k}\right)=\inprod{I_{\mathbb{A},j}}{I_{\mathbb{B},k}}=\left|\inprod{I_{\mathbb{A},j}}{I_{\mathbb{B},k}}\right|\leq\left\Vert I_{\mathbb{A},j}\right\Vert \left\Vert I_{\mathbb{B},k}\right\Vert =1\text{ .}
\end{eqnarray*}
Besides, by summation on the index $k$, one obtains:
\[
\sum_{k}p_{\mathbb{B}j,k}=\sum_{k}\tr\left(I_{\mathbb{A},j}I_{\mathbb{B},k}\right)=\tr\left(I_{\mathbb{A},j}\sum_{k}I_{\mathbb{B},k}\right)=\tr\left(I_{\mathbb{A},j}\right)=1\text{ .}
\]
The coefficients:
\begin{equation}
p_{\mathbb{B}j,k}=\tr\left(I_{\mathbb{A},j}I_{\mathbb{B},k}\right)=\inprod{I_{\mathbb{A},j}}{I_{\mathbb{B},k}}\label{eq:Transition_Prob}
\end{equation}
may be, therefore, interpreted as \emph{probabilities}. They are the
well-known \emph{transition probabilities} from pure states in the
initial space to pure states in the second one. If we now consider
the the projection $D_{\mathbb{A},\mathbb{B}}$, we have:
\[
D_{\mathbb{A},\mathbb{B}}=\sum_{j,k}p_{\mathbb{A},j}p_{\mathbb{B}j,k}\,I_{\mathbb{B},k}=\sum_{k}p_{\mathbb{AB},k}\,I_{\mathbb{B},k}
\]
where the coefficients:
\[
p_{\mathbb{AB},k}\ideq\sum_{j}p_{\mathbb{A},j}p_{\mathbb{B}j,k}
\]
satisfy again, as it easily verified, all the properties of a probability
distribution. The projection $D_{\mathbb{A},\mathbb{B}}$ has, therefore,
indeed the structure of a density observable.

Based on this observation and on \eqref{eq:Traccia_Val_Med}, the
expectation value of $B$ will be also given by:
\begin{equation}
\left\langle B\right\rangle _{\mathbb{A}}=\tr\left(D_{\mathbb{A},\mathbb{B}}B\right)\text{ .}\label{eq:Valore_Medio_Cond_2}
\end{equation}
Since the expectation value of the observable $B$ must be a well-defined
quantity, for consistency it must be:
\begin{equation}
\tr\left(D_{\mathbb{A}}B_{\mathbb{A}}\right)=\tr\left(D_{\mathbb{A},\mathbb{B}}B\right)\label{eq:Simmetria_Proiez}
\end{equation}
independently upon the choice of the complete space containing $B$.

To prove the validity of \eqref{eq:Simmetria_Proiez}, we observe
that, if $b_{k}$ is the spectral coefficient of $B$ relative to
the projector $I_{\mathbb{B},k}$, one has:
\begin{equation}
\tr\left(D_{\mathbb{A}}B_{\mathbb{A}}\right)=\sum_{j,k}p_{\mathbb{A},j}b_{k}\,\tr\left(I_{\mathbb{A},j}I_{\mathbb{B},k}I_{\mathbb{A},j}\right)\label{eq:Dim_Simm_Pr_1}
\end{equation}
and
\begin{equation}
\tr\left(D_{\mathbb{A},\mathbb{B}}B\right)=\sum_{j,k}p_{\mathbb{A},j}b_{k}\,\tr\left(I_{\mathbb{B},k}I_{\mathbb{A},j}I_{\mathbb{B},k}\right)\text{ .}\label{eq:Dim_Simm_Pr_2}
\end{equation}
But, according to the third hypothesis on the trace, it is obtained:
\[
\tr\left(I_{\mathbb{A},j}I_{\mathbb{B},k}I_{\mathbb{A},j}\right)=\tr\left(I_{\mathbb{A},j}\left(I_{\mathbb{B},k}I_{\mathbb{A},j}\right)\right)=\tr\left(\left(I_{\mathbb{B},k}I_{\mathbb{A},j}\right)I_{\mathbb{A},j}\right)=\tr\left(I_{\mathbb{B},k}I_{\mathbb{A},j}\right)
\]
and
\[
\tr\left(I_{\mathbb{B},k}I_{\mathbb{A},j}I_{\mathbb{B},k}\right)=\tr\left(\left(I_{\mathbb{B},k}I_{\mathbb{A},j}\right)I_{\mathbb{B},k}\right)=\tr\left(I_{\mathbb{B},k}\left(I_{\mathbb{B},k}I_{\mathbb{A},j}\right)\right)=\tr\left(I_{\mathbb{B},k}I_{\mathbb{A},j}\right)
\]
that, substituted in \eqref{eq:Dim_Simm_Pr_1} and in \eqref{eq:Dim_Simm_Pr_2},
prove the relation \eqref{eq:Simmetria_Proiez}, demonstrating the
physical consistency of the hypotheses taken on the trace of a pseudo-observable.

By these relations it also follows that:
\begin{equation}
\left\langle B\right\rangle _{\mathbb{A}}=\tr\left(D_{\mathbb{A}}B\right)\text{ .}\label{eq:Valore_Medio_Cond}
\end{equation}

Based on \eqref{eq:Valore_Medio_Cond}, it is therefore natural to
associate to a pseudo-observable $Z$, for a given state of a certain
physical system described by the density observable $D$, as its expectation
value $\left\langle Z\right\rangle $, the complex number:
\begin{equation}
\left\langle Z\right\rangle \ideq\tr\left(D\,Z\right)=\inprod DZ\text{ .}\label{eq:Valore_Medio_PO}
\end{equation}
According to such definition and to the properties of the trace, it's
easy to prove the following properties of the expectation value of
a pseudo-observable, generalizing those presented in subsection \ref{sec:Valori-medi}:
\begin{enumerate}
\item For two given pseudo-observables $Z_{1}$ and $Z$, the expectation
value of their sum is equal to the sum of their expectation values:\\
$\left\langle Z_{1}+Z_{2}\right\rangle =\left\langle Z_{1}\right\rangle +\left\langle Z_{2}\right\rangle $.
\item Given a pseudo-observable $Z$ and a complex constant $\gamma$, the
expectation value of their product is equal to the product of the
value of the complex constant by the expectation value of the pseudo-observable:\\
$\left\langle \gamma Z\right\rangle =\gamma\left\langle Z\right\rangle $.
\item For two given comparable observables $O_{1}$ and $O_{2}$, with $O_{1}\leq O_{2}$,
the expectation values follow the same order of the two observables,
so one has:\\
$\left\langle O_{1}\right\rangle \leq\left\langle O_{2}\right\rangle $.
\item The expectation value of the transposition of a pseudo-observable
$P$ is equal to the complex conjugate of the expectation value of
the pseudo-observable:\\
$\left\langle Z^{\dagger}\right\rangle =\left\langle Z\right\rangle ^{*}$.
\item The expectation value of constant $\gamma=\alpha+i\beta$ is equal
to the value of the constant itself:\\
if $\gamma=\alpha+i\beta$ then $\left\langle \gamma\right\rangle =\alpha+i\beta$.
\end{enumerate}
What proved in this section is coherent with the well-known \emph{Gleason's
theorem}\citep{Gleason1957}, of which what proved above plays an
analogous role in the framework of the algebra of the pseudo-observables,
even if the starting assumptions are different.

It is now important to clarify some points that can cause a misleading
comprehension about the role of the density observable. Effectively,
this observable does not correspond to an actually measured physical
quantity, but it is, instead, built up to incorporate every statistical
property of the physical system relative to a given choice of a complete
set of compatible observables. What happens then if, subsequently
to a certain maximum observation relative to a certain complete set
of compatible observables, one does a measurement of an observable
incompatible with some observable of the set? You are allowed to calculate
the expectation value of the new observable making use of the \eqref{eq:Valore_Medio_Cond_1},
but the measurement deeply alters the starting situation. In fact,
one switches from a description in terms of the density observable
relative to the starting complete set of compatible observables, to
another in which the density observable is relative to a new complete
set of compatible observables that \textbf{includes} the last measured
one and \textbf{excludes} those which are not compatible with it.
\textbf{This change does not, however, correspond to any physical
process}: the density observable is, in fact, a summary of the state
of the system, which incorporates all the statistical information
available to the observer. Such information refers to a description
of reality. But this description, as anticipated in subsection 2.1
of the first paper, represents the outcome of \textit{a posteriori}
process, based on the fundamental requirement of consistency. \textbf{When
to the information acquired through the maximum observation it is
added what obtained by the measurement of an observable incompatible
with some of the observable of the starting complete set, the picture
is no longer coherent and the new information deletes the older one,
incompatible with it}.

It must be, however, clear that the observation does alter the physical
state of the observed system, due to the unavoidable physical interaction
between the measurement probe and the system. Unlike the change in
the density observable, this is, instead, a physical process that
can give rise to physical effects. This point will be deepen in the
third paper.

\subsection{Deviations and uncertainty relations}

The difference between an observable $A$ and its expectation value
is given by its \emph{deviation} $\Delta A$:
\begin{equation}
\Delta A\ideq A-\left\langle A\right\rangle \text{ .}\label{eq:Def_Scarto}
\end{equation}
The deviation is an observable also, whose expectation value is equal
to zero:
\begin{equation}
\left\langle \Delta A\right\rangle =\left\langle A-\left\langle A\right\rangle \right\rangle =\left\langle A\right\rangle -\left\langle A\right\rangle =0\text{ .}\label{eq:Media_Scarto}
\end{equation}

The amount of variation or dispersion of an observable $A$ from its
expected value is usually quantified by introducing firstly the \emph{variance}
$\sigma_{A}^{2}$, defined as the expected value of the square of
its deviation: 
\begin{equation}
\sigma_{A}^{2}\ideq\left\langle \left(\Delta A\right)^{2}\right\rangle \label{eq:Def_Varianza}
\end{equation}
and then the \emph{standard deviation} $\sigma_{A}$:
\begin{equation}
\sigma_{A}\ideq\sqrt{\sigma_{A}^{2}}\text{ .}\label{eq:Def_DevSTD}
\end{equation}

A limitation above the values assumed for the standard deviations
of two incompatible observables is fixed by the well-known \emph{Heisenberg
uncertainty relations}. In order to derive in the new context these
relations, one can follows a procedure analogous to that used by Born
in the Appendix XXVI of his famous book\citep{Born1969}.

Let $A$ and $B$ be two pseudo-observables. For each value of a parameter
$\xi$, of a suitable physical dimension, consider the following pseudo-observable:
\begin{equation}
Z=\xi\Delta A+i\Delta B\label{eq:Comb_Scarti}
\end{equation}
According to the fifth property of transposition (see first paper),
one has:
\begin{eqnarray}
ZZ^{\dagger} & = & \left(\xi\Delta A+i\Delta B\right)\left(\xi\Delta A-i\Delta B\right)=\nonumber \\
 & = & \xi^{2}\left(\Delta A\right)^{2}-i\,\xi\left[\Delta A,\Delta B\right]+\left(\Delta B\right)^{2}\geq0\,.\label{eq:Comb_Scarti2}
\end{eqnarray}
According to the third property of the expectation value and by the
variance definition, one, then, obtains:
\begin{equation}
\sigma_{A}^{2}\,\xi^{2}-\left\langle i\left[A,B\right]\right\rangle \xi+\sigma_{B}^{2}\geq0\label{eq:Rel_Varianze}
\end{equation}
where the following immediate identity has also been exploited:
\begin{equation}
\left[\Delta A,\Delta B\right]=\left[A,B\right]\text{ .}\label{eq:Commutatore_Scarti}
\end{equation}
Since the \eqref{eq:Rel_Varianze} must hold for every value of the
parameter $\xi$, the left-hand side second degree trinomial cannot
have a positive discriminant. It, therefore, must be:
\[
\left|\left\langle i\left[A,B\right]\right\rangle \right|^{2}-4\,\sigma_{A}^{2}\sigma_{B}^{2}\leq0
\]
by which the Heisenberg uncertainty relation is obtained:
\begin{equation}
\sigma_{A}\,\sigma_{B}\geq\frac{1}{2}\left|\left\langle i\left[A,B\right]\right\rangle \right|\text{ .}\label{eq:Rel_Indeterminazione}
\end{equation}
The uncertainty relations states the impossibility of simultaneously
obtaining exact values from the measurement of two observable mutually
incompatible and thus represents a further aspect of the incompatibility
itself.

\subsection{Matrix elements and wave functions}

The introduction of an eigenstate space $\mathfrak{\mathbb{E}}=\sspan[\left(\left\{ \Psi_{j}\right\} \right)]$
allows one to write down some convenient expression for the dyadic
components of a pseudo-observable $P$. In fact, by making use of
the \eqref{eq:Coeff_Scomp_PO} and the \eqref{eq:Caratt_Vettori_Stato},
with reference to the decomposition (45) in the first paper:
\[
P=\sum_{j,k}\varpi_{jk}\,\Gamma_{jk}
\]
one has:
\begin{equation}
\varpi_{jk}=\inprod{\Gamma_{jk}}P=\tr\left(\Psi_{k}\Psi_{j}^{\dagger}P\right)=\tr\left(\Psi_{j}^{\dagger}P\Psi_{k}\right)=\inprod{\Psi_{j}}{P\Psi_{k}}\label{eq:Elementi_Matrice}
\end{equation}
by which, in particular, one obtains: 
\begin{equation}
\left\langle P\right\rangle _{j}\ideq\varpi_{jj}=\inprod{\Psi_{j}}{P\Psi_{j}}\label{eq:Valor_Medio_Stato_Puro}
\end{equation}
 that may be interpreted as the expectation value of $P$ in the pure
state of index $j$.

More generally, we can write down a new expression for the expectation
value of a whatever observable $O$. Observe, first of all, that each
probability $p_{j}$, defined by equation \eqref{eq:Def_Prob}, appearing
in the expression of the expectation value of $O$, may be interpreted
as the probability of \emph{occurrence} of the state $j$ after the
measurement. By exploiting the linearity of the inner product, according
to the equation \eqref{eq:Valore_Medio_PO} and to the \eqref{eq:Elementi_Matrice},
one has:
\begin{equation}
\left\langle O\right\rangle =\inprod DO=\sum_{j}p_{j}\inprod{I_{j}}O=\sum_{j}p_{j}\inprod{\Psi_{j}}{O\Psi_{j}}\label{eq:Valore_Medio_Prod_Sc_VS}
\end{equation}
that allows one to express the expectation value of an observable
in terms of inner products of state vectors.

By supposing to having identified the elements of an eigenstate basis
as simultaneous eigenvectors of a complete set of compatible observables
placed in a $n$-tuple $\mathbf{O}$, on the basis of what exposed
in the subsection \ref{sec:Spazio-autostati}, there will be a biunivocal
correspondence, apart from an equivalence, between each basis element
and a $n$-tuple $\mathbf{o}$ of eigenvalues of the observables in
$\mathbf{O}$.

According to the second spectral theorem, each element $\Phi\in\mathbb{E}=\sspan[\left(\left\{ \Psi_{\mathbf{o}}\right\} \right)]$
may, therefore, be decomposed in the form:
\begin{equation}
\Phi=\sum_{\mathbf{o}}\phi\left(\mathbf{o}\right)\,\Psi_{\mathbf{o}}\label{eq:Sviluppo_Funzione_Onda}
\end{equation}
where $\phi\left(\mathbf{o}\right)=\inprod{\Psi_{\mathbf{o}}}{\Phi}$,
by virtue of \ref{eq:Sviluppo_Autostati}, is a function of the $n$-tuple
$\mathbf{o}$ of eigenvalues, that will be called, with obvious reference
to the Schrödinger formalism, \emph{wave function of}\textbf{\textit{\emph{
$\Phi$}}}\emph{ relative to the basis}\textbf{\textit{\emph{ $\left\{ \Psi_{\mathbf{o}}\right\} $}}}.

If $\Phi_{1}$ and $\Phi_{2}$ are two elements of $\mathbb{E}$,
whose wave functions are, respectively, $\phi_{1}\left(\mathbf{o}\right)$
and $\phi_{2}\left(\mathbf{o}\right)$, due to the orthonormality
of the basis eigenstates and according to the properties of the inner
product, one has:
\begin{equation}
\inprod{\Phi_{1}}{\Phi_{2}}=\sum_{\mathbf{o}}\phi_{1}^{*}\left(\mathbf{o}\right)\phi_{2}\left(\mathbf{o}\right)\text{ .}\label{eq:Prod_Scal_Funzioni_Onda}
\end{equation}
Equation \eqref{eq:Prod_Scal_Funzioni_Onda} implies, in particular,
that the norm of an element $\Phi\in\mathfrak{\mathbb{E}}$ is given
by:
\begin{equation}
\left\Vert \Phi\right\Vert =\sqrt{\inprod{\Phi}{\Phi}}=\left(\sum_{\mathbf{o}}\left|\phi\left(\mathbf{o}\right)\right|^{2}\right)^{\nicefrac{1}{2}}\text{ .}\label{eq:Norma_Phi}
\end{equation}
According to the state superposition theorem, it can, therefore, be
affirmed that any of the nonzero elements of $\mathfrak{\mathbb{E}}$
such that for their wave function is finite the sum in the right-hand
side of the \eqref{eq:Norma_Phi} (\emph{square-summable function})
is proportional to a suitable eigenstate of some observable. In particular,
for an eigenstate $\Phi$ whose wave function is $\phi(\mathbf{o})$
it holds the \emph{normalization condition}:
\begin{equation}
\sum_{\mathbf{o}}\left|\phi\left(\mathbf{o}\right)\right|^{2}=1\text{ .}\label{eq:Condizione_Normalizzazione}
\end{equation}

Since each term in the summation in the left-hand side of \ref{eq:Condizione_Normalizzazione}
is non-negative, the condition is the closure relation typical of
probabilities. It is easy to verify, by making use of the \ref{eq:Caratt_Vettori_Stato},
the \ref{eq:Transition_Prob} and of the \ref{eq:Sviluppo_Funzione_Onda},
that:
\begin{equation}
p_{\mathbf{o}}\ideq\left|\phi\left(\mathbf{o}\right)\right|^{2}=\left|\inprod{\Phi}{\Psi_{\mathbf{o}}}\right|^{2}\label{eq:Trans_Prob_SV}
\end{equation}
is just the \emph{transition probability} from the pure state described
by the eigenstate $\Phi$ to that described by the eigenvector $\Psi_{\mathbf{o}}$,
i.e. that for which the measurement of each observable in the tuple
$\mathbf{O}$ has as outcome the corresponding value in the $n$-tuple
$\mathbf{o}$.

Suppose now that the state of a physical system is initially described
by the eigenstate $\Phi$ and then is performed a measurement of an
observable $O_{r}$ belonging to the complete set of observables in
the $n$-tuple $\mathbf{O}$. According to the \ref{eq:Valor_Medio_Stato_Puro},
the expectation value of the measurement will be:

\[
\left\langle O_{r}\right\rangle =\inprod{\Phi}{O_{r}\Phi}=\sum_{\mathbf{o},\mathbf{o'}}\phi^{*}\left(\mathbf{o}\right)\phi\left(\mathbf{o'}\right)\inprod{\Psi_{\mathbf{o}}}{O_{r}\,\Psi_{\mathbf{o'}}}
\]
where it was made use of the \ref{eq:Sviluppo_Funzione_Onda}. But
since $\Psi_{\mathbf{o'}}$ is an eigenvector of the observable $O_{r}$
and due to the orthonormality of the eigenstates, one has:

\begin{equation}
\left\langle O_{r}\right\rangle =\sum_{\mathbf{o},\mathbf{o'}}\phi^{*}\left(\mathbf{o}\right)\phi\left(\mathbf{o'}\right)o_{r}\,\delta_{\mathbf{o},\mathbf{o'}}=\sum_{\mathbf{o}}\left|\phi\left(\mathbf{o}\right)\right|^{2}o_{r}\text{ .}\label{eq:Val_Medio_Funz_Onda}
\end{equation}
By comparison of the \ref{eq:Val_Medio_Funz_Onda} with \ref{eq:Scomp_Val_Med},
one finally retrieves the famous \emph{Born Rule}\citep{Born1926},
that gives the probability $p\left(O_{r}=o\right)$ of obtaining the
outcome $o$ in the measurement of the observable $O_{r}$ of a physical
system initially described by the eigenstate $\Phi$:

\begin{equation}
p\left(O_{r}=o\right)=\sum_{\mathbf{o}}\left|\phi\left(\mathbf{o}\right)\right|^{2}\delta_{o_{r},o}\text{ .}\label{eq:Born_Rule}
\end{equation}

\section{Conclusions}

\subsection{A solution for the measurement problem}

On the basis of what stated in this paper, we can now try to propose
a solution for the measurement problem:
\begin{enumerate}
\item \textbf{State vectors}, in the framework of the algebra of the pseudo-observables,
\textbf{aren't fundamentals entities}, but only the elements of an
useful pseudo-observable set characterizing the \emph{pure state }resulting
by a previous observation of a complete set of compatible observables.
They cannot be defined individually, but only as a whole relative
to all the possible pure states of the physical system.
\item \textbf{The ``collapse'' of quantum states does not correspond to
any physical process}. The density observable, in fact, does not correspond
to an actually measured physical quantity, but it is, instead, a statistical
summary of the state of the system, which incorporates all the information
available to the observer. Such information refers to a description
of reality. But this description represents the outcome of \textit{a
posteriori} process, based on the fundamental requirement of consistency.
When to the information acquired through the maximum observation it
is added what obtained by the measurement of an observable incompatible
with some of the observable of the starting complete set, the picture
is no longer coherent and\textbf{ the new information deletes the
older one incompatible with it}. On this question This last sentence
may seem puzzling and requires a deeper discussion that will be presented
in the next subsection.
\item We agree with Fuchs' quote: ``\textbf{QUANTUM STATES DO NOT EXIST}''.
The quantum state of a physical system is \textbf{undefined} till
the moment it is performed a measurement of some observables and it
is \textbf{not completely defined} till the moment it is measured
a complete set of compatible observable. However, even then, the density
observable characterizing the quantum state depends on the assignment
of a probability distribution, which can be done only on the basis
of the degree of belief of the \textbf{observer} in the occurrence
of the events associated to the pairwise orthogonal primitive projectors
of a basis, according to a given set of information.
\item \textbf{Quantum states are}, besides,\textbf{ observer-relative}:
different \emph{independent} observers (i.e. not communicating each
with other) may have different sets of information, deriving by different
sets of observational outcomes. They will, therefore, assume different
density observables, till to the extreme case in which for an observer
the system is in a pure state whereas for the other is in a mixed
one. The \emph{relational} view of Rovelli\citep{Rovelli1996} is
so necessarily integrated in our vision.
\item Whereas the ``perceptions'' of two independent observers may disagree,
in the moment they communicate each with other, they acquire a new
bunch of information and, consequently, they are forced to assume
a new common description of the system. \textbf{The exchange of information}
(communication) therefore \textbf{causes a ``collapse'' not only
of the quantum state of the physical system measured but also of the
``cognitive'' states of the two observers}. This apparently paradoxical
situation rises from the fact that each observer is a subject with
respect to himself and an object with respect to the other.
\item \textbf{The exchange of information between the two observers may},
thus, \textbf{be considered} as a process of reciprocal observation,
or, better, \textbf{as a measurement performed by the meta-observer
generated by the two observers of the physical system under study}.
Besides, since also the instrumental setups used by each of the the
observers to perform their measurements, according to the conclusive
discussion made in the first paper, may also be regarded as ``passive''
observers, \textbf{one may always think to measurement as a flow of
information from the physical system to the observer, trough the measurement
instrument, giving rise to a logical higher order meta-observer}.
\item \textbf{Quantum states are}, then, \textbf{properties of the meta-observers}
generated by the information exchanges associated to the measurements.
They are, therefore, ``relational'' entities\citep{Rovelli1996}.
\item Quantum states evolves through the time in a continuous manner, according
the time evolution equations, because during this process \textbf{no
information exchange occurs and no new meta-observer arises or changes}.
We will return to this point in the next paper.
\end{enumerate}
In order to clarify the last three points, we may think to the example
of the detection of the two-slit interference pattern generated by
single electrons, a remarkable experimental example of which is in
\citep{Merli1976}. For the electrons can generate the interference
pattern, it is necessary that they have the same linear momentum,
that must be therefore measured \textbf{before} each electron reaches
the slits. The position of the electron, \textbf{after} it has traversed
the slits, is then detected on a viewing screen. The point is that
since moment and position measurements are incompatible one with each
other, the initial meta-observer (experimenter$+$instrumentation
for moment measurement) is \textbf{different} from the final one (experimenter$+$viewing
screen), and so it will be also for relative quantum states!

What above outlined, in my opinion, remove every ambiguity and obscurity
in the quantum process of measurement.

The picture, indeed, may be difficult to be accepted, but it is based
on a deep analysis of the involved processes, it conforms to experimental
evidence and it is a logical consequence of the hypotheses assumed
to build the theory of measurement in the framework of the algebra
of the pseudo-observables. So, I think, it is \textbf{the only} interpretation
possible in this context.

One could object that quantum mechanics may be formulated in other
different forms, but the construction of the algebra of the pseudo-observables
is such to leave little space, if any, to nonequivalent alternatives.
Moreover even the apparently equivalent ones, like that of Dirac-von
Neumann, suffer for the introduction of unneeded extra concepts that
cause ambiguities in the interpretation: the assumption of quantum
states as the basis concept was, in my views, like putting the cart
before the horse, since observations \textbf{precede} and \textbf{not
follow} quantum states definition!

A remarkable final consideration is that Relational Quantum Mechanics\citep{Rovelli1996,sep-qm-relational}
and QBism\citep{Fuchs2010,Fuchs2011,Fuchs2013} may both be considered
as two compatible \emph{interpretations} of the formulation of Quantum
Mechanics in term of the algebra of the pseudo-observables outlined
in this and my previous paper.

\subsection{What is real?}

In this closing subsection, we want discuss about the deeper conceptual
consequences of picture of the reality that results by our theory.

First of all, if it is assumed the validity of the new formulation
and its embedded vision of the world, quantum states and, consequently,
wave functions cannot be ontologically real, due to their relational
nature. Therefore, de Broglie-Bohm theory, but also many world ones,
based on the concept of an ontologically real universal wave function,
are incompatibles with my vision. Moreover, since the quantum state
``collapse'' is a logical process and not a physical one, the theories
of objective collapse would have also to be ruled out.

The final question we want to analyze is what picture of the reality
emerges from our vision. The starting point has mandatory to be the
so called EPR argument\citep{EPR}, conceived as an attack against
the description of measurements according the Copenhagen interpretation
and a criticism of the idea that Quantum Mechanics could be a complete
description of reality. Instead of a detailed exposition of the argument,
for which a convenient reference is made to the paper of Smerlak and
Rovelli\citep{Smerlak2007}, here we want simply to show as the failure
of the argument, experimentally demonstrated by Aspect \textit{et
al.}\citep{Aspect1982a,Aspect1982b}, gives a new insight about reality.

The argument was based on the assumption of the three fundamental
concepts of realism, locality and separability. According to Einstein,
\emph{realism} is meant as the assumption\citep{Einstein1950} that:
\begin{quotation}
\textquotedblleft There exists a physical reality independent of substantiation
and perception\textquotedblright{}
\end{quotation}
The concept of \emph{separability} is also well illustrated by Einstein's
words\citep{Einstein1948}:
\begin{quotation}
``Without such an assumption of the mutually independent existence
\ldots{} of spatially distant things, as assumption which originates
in everyday thought, physical thought in the sense familiar to us
would not be possible. Nor does one see how physical laws could be
formulated and tested without such a clean separation.''
\end{quotation}
The enunciation of the \emph{locality} principle is, implicitly, reported
in the EPR paper\citep{EPR}:
\begin{quotation}
``since at the time of measurement the two systems no longer interact,
no real change can take place in the second system in consequence
of anything that may be done to the first system.''
\end{quotation}
All of the three statements sound obvious to plain common sense, but,
this, following the argumentation of the EPR paper, would imply that
Quantum Mechanics is not a \emph{complete} theory, i.e., according
to the Einstein's definition, that \textbf{not} every element in the
physical reality has a counterpart in the theory.

John Bell, in 1966, showed that no complete local realistic theory
can be compatible with Quantum Mechanics\citep{Bell1966}. The result
of Bell, however, left open the way to the existence of complete non-local
hidden variable theories (such that of Bohm). But giving up locality
is even worse than sacrificing realism. In fact non-locality implies
that an event here may be \textbf{instantly} the cause of an effect
in a distant point, and due to relativity, this would imply that for
some observer the effect will precede the cause! \textbf{So non-locality,
at a fundamental level, destroys causality and so also Physics!}

The Kochen-Specker theorem\citep{Kochen1967} gave what may considered
a blow of grace to realism. In fact, they prove rigorously the impossibility
of a value assignment to an observable before the measurement act,
independently by the compatible observables measured with it.

The waiver to realism is a compelling issue. In a conversation with
A. Pais, Einstein, to this regards, suddenly asked if he thought that
the moon existed only when he looked at it\citep[p. 907]{Pais:1979vn}.
The point is that the whole REALITY DOES NOT EXIST, in the sense that
it is only an emerging property, an \textit{a posteriori} reconstruction,
resulting by the exchange of information in a closed network of observer
(that thus give rise to a single meta-observer). It is an hard truth
to accept, but the logical and mathematical analyses and the experimental
results points only in this direction!

Like in Plato's Cave, we lived exchanging shadows for the truth\ldots{}
\begin{quotation}
«\textit{Stat rosa pristina nomine, nomina nuda tenemus}»

(The rose, which was, {[}now{]} exists only in the name, we only possess
bare names)\citep{Nome_Rosa}.
\end{quotation}

\appendix

\section{Two properties of inner product of primitive projectors\label{sec:Appendice}}

\subsection{If $I$ and $J$ are two primitive projectors and $\protect\inprod JI=0$,
then $I$ and $J$ are compatibles and orthogonal\label{subsec:Lemma_1}}

In fact one has:
\[
0=\inprod JI=\tr(J\,I)=\tr(J\,I^{2})=\tr(I\,J\,I)\text{ .}
\]
But for the projection $I\,J\,I$, by making use of \eqref{eq:Coeff_Scomp_PO},
it holds:
\[
I\,J\,I=\inprod{I\,J\,I}II=\tr(I\,J\,I)\,I=0\text{ .}
\]
So that it results:
\[
0=I\,J\,I=I\,J\,J\,I=(I\,J)(I\,J)^{\dagger}
\]
that, by virtue of the sixth property of the Hermitian transposition
presented in the first paper, implies:
\[
I\,J=0
\]
and therefore:
\[
0=(I\,J)^{\dagger}=J\,I\text{ .}
\]
For these last two relations one, finally, has:
\begin{eqnarray*}
I\,J=J\,I=0\\
\left[I,J\right]=0
\end{eqnarray*}
so that the initial assertion is proved.

\subsection{If $I$ and $J$ are two primitive projectors and $\protect\inprod JI=1$,
then $J=I$ \label{subsec:Lemma_2}}

As seen for the lemma \eqref{subsec:Lemma_1}, the hypothesis implies
that it results:
\[
I\,J\,I=\inprod{I\,J\,I}II=I
\]
so that one has:
\[
0=I-I\,J\,I=I(1-J)I
\]
where $1-J$ is the complementary projector to $J$. From this, it
follows:
\[
0=I-I\,J\,I=I(1-J)I=I(1-J)^{2}I=\left(I(1-J)\right)\left(I(1-J)\right)^{\dagger}
\]
that implies:
\[
I(1-J)=0\;\Rightarrow\;I\,J=I
\]
and then, similarly to what observed in \ref{subsec:Lemma_1}:
\[
\left[I,1-J\right]=0\;\Rightarrow\;\left[I,J\right]=0\text{ .}
\]
Therefore one has:
\[
I\,J=J\,I=I\text{ .}
\]
If we, now, put:
\[
K\ideq J-I\text{ ,}
\]
it results:
\[
K^{2}=J-J\,I-I\,J+I=J-I=K\text{ ,}
\]
so that $K$ is a projector. But since it results:
\[
J=I+K
\]
and $J$ is primitive, then it have to be $K=0$ and therefore:
\[
J=I
\]
that is what we wanted to proof. 

\ack{}{}

I want to kindly thank David Edwards for the useful discussions and
suggestions to improve this paper.\bibliographystyle{unsrt}
\addcontentsline{toc}{section}{\refname}\bibliography{../Bibliography}

\end{document}